\documentclass[preprint,superscriptaddress,noshowpacs]{revtex4-1}%
\usepackage{graphicx}
\usepackage{subfigure}
\usepackage{amsmath}
\usepackage{mathtools}
\usepackage[verbose]{placeins}
\usepackage[hang,flushmargin]{footmisc}
\usepackage{amssymb}%
\usepackage{natbib}
\usepackage{xcolor}
\usepackage{epsfig}
\usepackage{multirow}
\usepackage{nicematrix,booktabs}
\usepackage{dcolumn}
\usepackage{grffile}
\usepackage{float}
\usepackage[utf8]{inputenc}
\usepackage[T1]{fontenc}
\usepackage{gensymb}
\usepackage{cleveref}
\usepackage{tabularx}
\usepackage[labelsep=quad,indention=10pt]{subfig}
\usepackage[skip=0pt]{caption} % default 10pt
\usepackage[nodisplayskipstretch]{setspace}

\begin{document}
%\title{Adsorption and Oxidation of Formaldehyde molecule on P-doped CeO$_{2}$(111) facet using first-principles calculations}
\title{
Synergistic Effects of Phosphorus Doping and Oxygen Vacancies on Formaldehyde Oxidation over CeO$_2$(111): A First Principles Investigation}

\author{Tarek Ayadi}
\affiliation{Université de Lorraine, CNRS, LEMTA, F-54000 Nancy, France}%

\author{Mourad Debbichi}
\affiliation
{Université de Monastir, Faculté des Sciences de Monastir, Laboratoire de la matière condensée et nanosciences LR11ES40, 5019
Monastir, Tunisia}

\author{Michael Badawi}
\affiliation
{Laboratoire Lorrain de Chimie Moléculaire, Université de Lorraine, CNRS, Nancy F-54000, France}

\author{Fabien Pascale}
\affiliation
{Université de Lorraine, CNRS, LEMTA, F-54000 Nancy, France}

\author{Adel Mesbah}
\affiliation
{Université Lyon 1, CNRS, IRCELYON, UMR 5256, Villeurbanne, France}

\author{Sébastien Leb\`egue}
\affiliation
{Université de Lorraine, CNRS, LEMTA, F-54000 Nancy, France}

\begin{abstract}
 Using a combination of static and dynamic density functional theory simulations, we systematically investigated how phosphorus doping and oxygen vacancies on the CeO$_2$(111) surface influence the oxidation mechanisms of formaldehyde (HCHO). Our results reveal that P cations (P$^{5+}$) substitutionally replace Ce$^{4+}$ in the lattice, forming Ce$-$O$-$P bonds that reduce the band gap (from 2.26 eV to 2.09 eV) and generate localized Ce$^{3+}$ states through charge redistribution. This synergistic effect of P doping combined with oxygen vacancy strengthens HCHO adsorption by decreasing the adsorption energy from -0.62 eV on pristine CeO$_2$(111) to -2.65 eV on the defective P-doped surface.
 Importantly, P doping lowers the C$-$H bond cleavage barrier by 0.84 eV relative to pristine CeO$_2$(111), accelerating formaldehyde oxidation on the defective surface. In addition, the rapid desorption of CO$_2$ and H$_2$O ($\tau \sim 0.59 s$ at 300 K) indicates weak product-surface interactions, which favor efficient catalyst regeneration during continuous operation. These findings highlight P-doped CeO$_2$(111) as a promising system for low-temperature HCHO oxidation and provide insights into the design of ceria-based catalytic materials.
\end{abstract}
\maketitle

%\section*{Keywords}

%DFT, AIMD, NEB, Heterogeneous catalysis, Formaldehyde, CeO$_2$, surface, Phosphorus doping,

\section{Introduction}

Formaldehyde (HCHO), a major toxic indoor pollutant emitted from building materials, is well recognized as being highly harmful to human health {\color{blue}\cite{Chen2017,Liu2023,He2021,LIU2023110080,ZHOU2025137831}}. Owing to its serious health risks, the development of effective removal methods is essential, particularly under low- and room-temperature conditions. Although various methods, such as photocatalytic oxidation, adsorption, and plasma techniques, have been explored, catalytic oxidation stands out as a highly efficient approach for the complete conversion of formaldehyde into harmless CO$_2$ and H$_2$O. Consequently, extensive research efforts have been devoted to the development of efficient catalysts for this process. Among them, metal oxides including SiO$_2$, MnO$_2$, Mn$_3$O$_4$, Co$_3$O$_4$, TiO$_2$, and CeO$_2$ have been widely explored {\color{blue}\cite{QI2021106293,DENG2019540}}. Among these materials, CeO$_2$ has emerged as a particularly promising catalytic system owing to its outstanding redox properties, which are associated with the facile and reversible Ce$^{4+}$/Ce$^{3+}$ redox couple, high oxygen storage and activation capacity, good thermal stability, and relatively low cost.

To facilitate the development of more efficient low-temperature catalysts for HCHO removal, the oxidation mechanisms of HCHO on ceria surfaces have been extensively investigated. A density functional theory (DFT) study by Tang et $al.$ {\color{blue}\cite{TENG201068}} on the CeO$_2$(111) surface revealed that HCHO desorption is favored on the stoichiometric surface due to the high energy barriers associated with C–H bond cleavage. In contrast, the presence of oxygen vacancies significantly lowers these barriers, thereby promoting the formation of CO and H$_2$ {\color{blue}\cite{BREYSSE197354,TENG201068}}. Furthermore, O$_2$ adsorption, particularly at vacancy sites, substantially enhances the oxidation efficiency by replenishing active oxygen species {\color{blue}\cite{TENG201068}}. Pure CeO$_2$ catalysts generally exhibit limited activity toward HCHO oxidation. However, their catalytic performance can be markedly enhanced through morphology control and by incorporating other metal oxides or noble metals. For example, complete conversion of HCHO into CO$_2$ and H$_2$O has been achieved via catalytic oxidation over Pt/MnO$_x$–CeO$_2$ catalysts at around room temperature, as well as over Au/CeO$_2$ catalysts at approximately 75 °C {\color{blue}\cite{TANG2008115,LIU2012467}}. Qin et $al.$ {\color{blue}\cite{D0CY01894E}} highlighted the potential of Ru-based catalysts for indoor air purification by investigating the catalytic oxidation of formaldehyde (HCHO) over highly efficient Ru/CeO$_2$ and Ru/Al$_2$O$_3$ systems. Their results demonstrated that Ru/CeO$_2$ outperformed Ru/Al$_2$O$_3$, achieving complete HCHO conversion at 90 °C. In a complementary theoretical study, Zhang et $al.$ {\color{blue}\cite{ZHANG2020121693}}  employed DFT+U calculations to elucidate the reaction mechanisms of HCHO oxidation, with particular emphasis on the role of oxygen vacancies and the effectiveness of Co doping in enhancing catalytic activity. These insights provide guidance for the rational design of cost-effective and environmentally friendly catalysts capable of operating at room temperature.

So far, research efforts have predominantly focused on noble metals (such as Pt, Au, Rh, and Ag) and transition metal oxides for HCHO oxidation at room temperature. Although noble metals generally exhibit superior catalytic activity, their high cost and limited availability significantly restrict their large-scale and practical applications. It has been suggested that heteroatoms, particularly nitrogen, can significantly enhance catalytic activity in the selective catalytic reduction of nitrogen oxides for automotive applications, and also exhibit a notable capability for the degradation of various organic pollutants {\color{blue}\cite{NING2024133282,acsomega.3c01305,D0NJ03270K, HAN2022132154}}. Phosphorus shares the same number of valence electrons as nitrogen; however, its larger atomic radius and stronger electron-donating ability render it more effective for oxygen reduction reactions {\color{blue}\cite{HAN2022132154}}. Moreover, the high abundance and cost-effectiveness of phosphorus make it an attractive candidate for large-scale industrial applications. Phosphorus doping (P-doping) has also been shown to enhance electrochemical performance by increasing the density of free charge carriers, facilitating electron and ion transport, and generating additional active surface sites {\color{blue}\cite{YOU201847}}.

Over the past two decades, only a limited number of studies have highlighted the significant role of phosphorus doping in regulating the activity, selectivity, and stability of metal oxide–based catalysts. In TiO$_2$, phosphorus can be incorporated in both cationic (P$^{5+}$) and anionic (P$^{3-}$) forms: P$^{5+}$ doping has been reported to induce a blue shift and increase the density of free charge carriers, whereas P$^{3-}$ species promote faster charge recombination, thereby influencing the photocatalytic activity {\color{blue}\cite{cm504734a}}.
Cerium phosphate–based materials exhibit excellent thermal and chemical stability, along with suitable acidity, making them advantageous for catalytic applications {\color{blue}\cite{acs.est.2c00942, acs.jpcc.0c03649,acs.jpcc.5b07734}}. In particular, CePO$_4$ and samarium (Sm)-doped CePO$_4$ have emerged as promising catalysts for the oxygen reduction reaction (ORR), owing to their unique ability to simultaneously provide Ce$^{3+}$ ions and oxygen vacancies within the crystal structure {\color{blue}\cite{GUPTA2022100947}}. Furthermore, Pt/CePO$_4$ catalysts have demonstrated high activity and selectivity in the water-gas shift (WGS) reaction {\color{blue}\cite{ C8TA04603D}}. More recently, Dai et $al.$ {\color{blue}\cite{acs.est.4c04436}} reported that phosphate-based CePO$_4$ catalysts exhibit superior thermal and chemical stability, along with remarkable efficiency in the catalytic hydrolysis–oxidation of halogenated volatile organic compounds (HVOCs).

Although pure CeO$_2$ exhibits limited catalytic activity due to its weak surface acidity, it has been demonstrated that phosphorus incorporation can significantly enhance catalytic activity, thermal stability, and ion-exchange properties, all of which are crucial for nitrogen oxide removal {\color{blue}\cite{YOU201847}}. More recently, $Kaili ~et ~al.$ {\color{blue}\cite{GONG2020144314}} successfully synthesized two-dimensional (2D) P-CeO$_2$ hybrids via a bio-assisted calcination method. Their results showed that these hybrids act as highly effective flame retardants for epoxy resins, markedly improving thermal stability and substantially reducing fire hazards by suppressing heat release, smoke emission, and carbon monoxide (CO) production. In addition, the formation of a robust and protective char layer further highlights the potential of P-CeO$_2$ hybrids as promising additives for enhancing the fire safety of polymer-based materials.

Collectively, these studies indicate that phosphorus modification can profoundly alter the physicochemical properties of cerium-based materials, highlighting its potential for tuning catalytic performance in oxidation reactions under mild conditions. Non-metal cation doping is an effective strategy to tailor the electronic structure of CeO$_2$, enabling band-gap narrowing and the formation of a Ce$^{4+}$/O$^{2-}$ charge imbalance that promotes electron transfer and suppresses charge recombination. Nevertheless, the impact of phosphorus (P) incorporation in ceria-supported catalysts has not yet been systematically investigated. To address this knowledge gap, we systematically investigate, using density functional theory (DFT) calculations at both 0 $K$ and finite temperatures, the interaction of phosphorus species supported on ceria (CeO$_2$) and their resulting effects on HCHO oxidation performance and reaction pathways.

\section{Computational details and methods}

In this work, all the calculations were based on the spin-polarized density functional
theory (DFT) using the Vienna {\it ab initio} simulation package
(VASP) {\color{blue}\cite{KRESSE1996}}. The electron-exchange correlation functional
was modeled via the generalized gradient approximation (GGA) using the
Perdew-Burke-Ernzerhof (PBE) functional{\color{blue}\cite{GGA1996}}.
The projector-augmented-wave (PAW) was used to describe the interaction between ions
and valence electrons.
Calculations were performed with a kinetic energy cutoff of 520 eV for the plane-wave basis set. We used 3$\times$ 3 $\times$1 and 9 $\times$ 9 $\times$1 Monkhorst-Pack grids for sampling the Brillouin zone during the geometry optimization (SCF) and for electronic properties calculations, respectively.
To account for the strong electronic correlations of the Cerium (Ce) 4f states,
the DFT+U method was applied with an effective U (U$_{eff}$)
value of 5.0 eV {\color{blue}\cite{PhysRevB.57.1505}}.

The CeO$_2$(111) surface was modeled using a (3$\times$3) surface supercell (a = 11.43 \AA~ and b= 11.49 \AA), with a vacuum spacing of 15 \AA~ along the c-direction to eliminate spurious interactions between periodic images. The (111) surface of CeO$_2$ was selected for the simulations as it is recognized as the most thermodynamically stable surface {\color{blue}\cite{D5CP01283J}}. To describe the Van der Waals (vdW) interactions between adsorbed molecules and the surface,
we integrated Grimme's DFT-D3 method {\color{blue}\cite{Grimme10}}. The slab model consisted of six atomic layers, of which the bottom three were fixed at their bulk-truncated positions, while the three upper layers were fully relaxed during geometry optimization. All remaining atoms, as well as all adsorbates, were fully relaxed until the Hellmann–Feynman forces acting on each atom were reduced to below 0.025 eV.\AA$^{-1}$.\\
Bader charge analysis was carried out to study the charge distribution
and transfer {\color{blue}\cite{Bader}}. The value of the Bader charge corresponds to the number of valence electrons of the neutral atom minus its population of Bader electrons.
The climbing image nudged elastic band (CI-NEB) method as implemented
in VASP {\color{blue}\cite{neb}} was used to identify the transition states of chemical reactions.\\
Thermal stability was evaluated at 300 K via AIMD simulations, using the electronic parameters described above for static DFT, in the NVT ensemble
for a total duration of 100 ps, utilizing a 1 fs time step. The temperature is regulated with a Nos\'e$-$Hoover thermostat {\color{blue}\cite{Nose1984, Hoover1985}}.  To avoid fictitious molecular splitting, the mass of the hydrogen atom is increased to 3 which corresponds
to the atomic mass of tritium{\color{blue}\cite{Ayadi2022}}.\\
The charge density diﬀerences of system AB were evaluated using the formula:
\begin{equation}
 \Delta\rho=\rho(AB)-\rho(A)-\rho(B)
\end{equation}
The formation energy, E$_{f}$, of the P-doped CeO$_{2}$(111) was calculated by:
\begin{equation}
 E_{f}=E_{PCeO_{2}}-E_{P}-E_{CeO_{2}}+E_{Ce}
 \label{efor}
\end{equation}
where E$_{PCeO_{2}}$ is the total energy of P-doped system, E$_{CeO_{2}}$ is
the energy of the clean CeO$_{2}$ surface,  and E$_{P}$ and E$_{Ce}$ are the
total energy of P and Ce atoms, respectively in their pure solid form.
The adsorption energy of the molecule (M) on the P-doped CeO$_{2}$ surface was
calculated as:
\begin{equation}
 E_{ad}(M)=E_{M/PCeO_{2}}-E_{M}-E_{PCeO_{2}}
\end{equation}
where E$_{M/PCeO_{2}}$ stands for the total energy of the adsorbed system,
E$_{PCeO_{2}}$ for the substrate energy, and E$_{M}$ for the gas molecule energy.\\
 In order to study the thermodynamics of O-vacancy formation,
we defined vacancy formation energy as:
\begin{equation}
 E_{f}(V_{O})=E_{def}-E_{PCeO_{2}}+\mu_{O}
\end{equation}
where E$_{def}$ is the energy of the defective surface and $\mu_{O}$ is the oxygen chemical potential.

To maintain consistency with standard catalysis literature and focus on the primary descriptor governing the catalyst's reactivity, we restricted our calculations to the standard O-rich limit ($\mu_{\text{O}} = \frac{1}{2}E_{\text{O}_2}$). E$_{O_{2}}$ is the energy of the O$_{2}$ molecule.
It should be noted that all the calculated values of E$_{f}(V_{O})$ in this work are referenced to gas-phase O$_{2}$ and carry the systematic overestimation of O$_{2}$ binding energy inherent to PBE{\color{blue}\cite{ PhysRevMaterials7065403}}.

Transition states (TSs) were determined using the climbing
image nudged elastic band method (CI-NEB){\color{blue}\cite{10.10631.1329672}} and veriﬁed by
vibrational analysis with only one imaginary frequency. Seven images are used, and the structures of all images are relaxed with a force convergence criterion set up to 10$^{-3}$ eV/\AA{}.
The energy barrier (Ea) was determined as the energy diﬀerence between the corresponding transition and initial states.
\section{Results and discussion}

\subsection{Single P-Doped CeO$_{2}$(111)}

\begin{figure}[H]
   \begin{center}
       \includegraphics[angle=0,width=0.8\linewidth,clip]{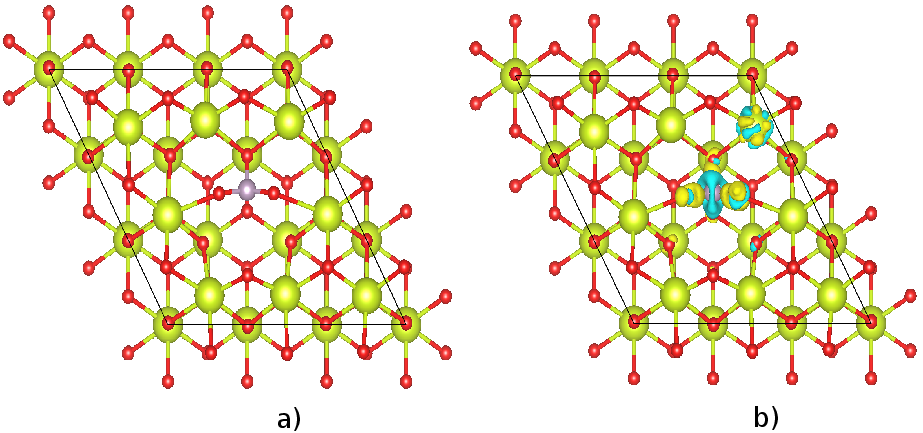}
     \caption{
(a) Top view of the optimized ($3 \times 3$) P-doped CeO$_2$(111) surface.
(b) Charge-density difference upon phosphorus doping on the CeO$_2$(111) surface. Yellow and blue regions represent electron accumulation and electron depletion, respectively. Cerium, oxygen, and phosphorus atoms are depicted as yellow, red, and purple spheres, respectively.
}
    \label{struc}
    \end{center}
\end{figure}

To investigate the structural stability of the P atom doping on the CeO$_2$(111) surface, all possible substitutional and adsorption sites were systematically examined. Our results indicate that the most energetically favorable configuration corresponds to a P atom substituting for a Ce atom, as illustrated in Fig. {\color{blue}\ref{struc}}. As shown in Fig. {\color{blue}\ref{struc}a}, the phosphorus atom is coordinated to two surface oxygen atoms with bond lengths of 1.50 \AA~ and to one subsurface oxygen atom with a bond length of 1.55~\AA. These P$-$O bonds are significantly shorter than the corresponding Ce$-$O bonds, leading to a localized structural distortion in the vicinity of the P dopant while preserving the overall crystal structure of the pristine surfaces. As a consequence, the coordination number of the two surface oxygen atoms is reduced from three to two. The calculated P$-$O vibrational frequencies at 1107.34 and 990.17 cm$^{-1}$ can be assigned to P$-$O and P$=$O stretching modes {\color{blue}\cite{GONG2020144314}}, respectively, and show good agreement with experimental Fourier-transform infrared (FT-IR) spectroscopy data. These results are consistent with the experimental observations reported by $Kaili ~et ~al.$ {\color{blue}\cite{GONG2020144314}}, as well as with other previously published studies {\color{blue}\cite{chem.201502170}} .
%%%%%%%%%%%%%%%%%%%%%%%%%%%%%%%%%%%%%%%%%%
%%%%%%%%%%%%%%%%%%%%%

Also, the formation energy value obtained from Eq.\ref{efor} is positive (3.32 eV),
indicating that an energetic input is required to achieve doping of CeO$_{2}$ surface with P. This value is in the same order of magnitude with M-doped
CeO$_{2}$(M=Mn, Co, Cu, Zr, Re and Mo){\color{blue}\cite{LIU2024119544,D3NR05950B}}. Although the positive formation energy (3.32 eV) indicates that substitutional P incorporation at the Ce site is thermodynamically metastable relative to separated CeO$_{2}$ and CePO$_{4}$ phases under equilibrium conditions, such configurations are experimentally accessible under non-equilibrium synthesis routes, including high-temperature calcination and solvothermal processing where kinetic stabilization of metastable dopant states is well established in doped oxide systems{\color{blue}\cite{acs.est.4c04436, GONG2020144314, acs.chemrev.5b00603}}.

The Bader charge analysis reveals a net charge of +3.20$\lvert e \rvert$ for the phosphorus atom. This value is highly comparable to the Bader chargres in reference P$_{2}$O$_{5}$ {\color{blue}\cite{acsanm3c02656} }and CePO$_{4}$ system, which is consistent with the heavily oxidized P$^{5+}$ like phosphate-type coordination environment. \\
Moreover, the three-dimensional charge-density difference of the P-doped CeO$_{2}$(111)
system was calculated and shown in Fig.{\color{blue}\ref{struc}b}.
The analysis reveals that the P atom, which possesses one more valence electron than Ce,
induces charge transfer towards neighboring atoms, resulting in the reduction
of one Ce$^{4+}$ ion to Ce$^{3+}$ in the vicinity of the P dopant.

\begin{figure}[H]
   \begin{center}

       \includegraphics[angle=0,width=0.8\linewidth,clip]{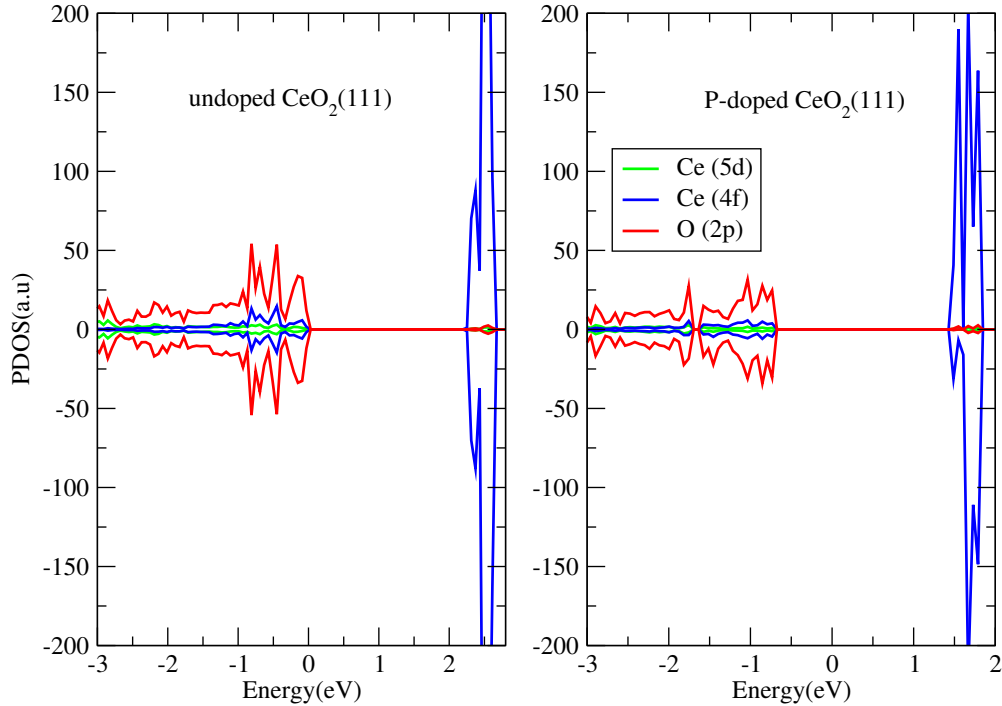}
     \caption{ The projected density of states (PDOS) of the undoped (left panel) and P-doped (right panel)
     CeO$_{2}$(111) surface.}
    \label{dos}
    \end{center}
\end{figure}

To further elucidate the interaction between the P dopant and the CeO$_2$(111) surface, we investigated the electronic structure of the system. Figure {\color{blue}\ref{dos}} presents the projected density of states (PDOS) of both pristine and P-doped CeO$_2$(111). As shown in Fig. {\color{blue}\ref{dos}}, pristine CeO$_2$(111) is a nonmagnetic semiconductor with a band gap of 2.26 eV, originating from the separation between the O 2p valence band and the empty Ce 4f conduction band. Upon P doping, the PDOS of the doped system exhibits distinct shifts compared to the pristine one, indicating that P doping leads to significant modifications in the electronic structure. Notably, the spin-up and spin-down electronic states are nearly degenerate, indicating a weak spin polarization of the system. In particular, the spin-up channel exhibits a charge transfer from the dopant to the surrounding lattice, leading to the reduction of a neighboring Ce$^{4+}$ cation to Ce$^{3+}$. This charge redistribution gives rise to localized Ce 4f states within the band gap, resulting in a band gap narrowing to 2.09 eV.

%##############################

%##############################
\subsection{Formaldehyde adsorption}
After establishing the structural, electronic, and charge-transfer properties of the pristine and P-doped CeO$_2$(111) surfaces, we now turn to the adsorption behavior of formaldehyde (HCHO) on the P-doped surface. Since adsorption constitutes the first and often rate-determining step in heterogeneous catalysis, understanding the interaction between HCHO and the modified ceria surface is essential for elucidating the origin of the enhanced catalytic activity induced by phosphorus doping.

%##############################
\begin{figure}[H]
   \begin{center}
       \includegraphics[angle=0,width=0.65\linewidth,clip]{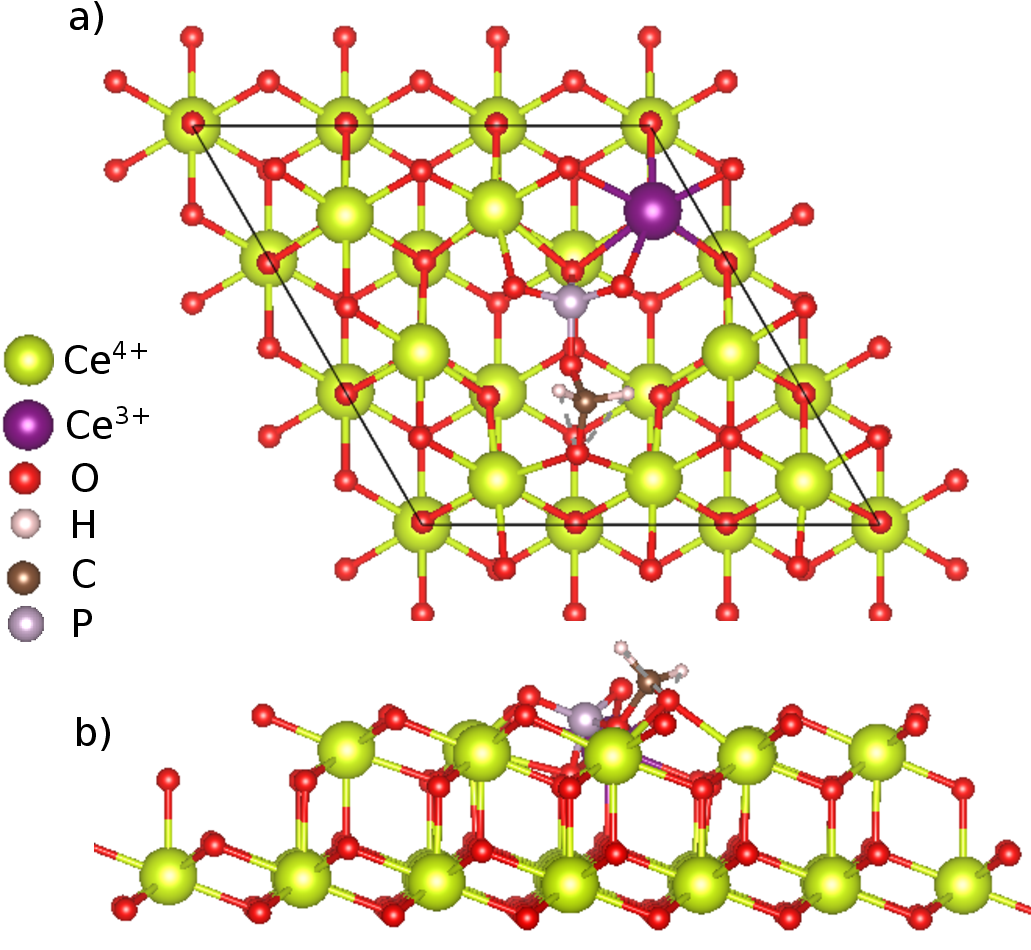}
     \caption{a)Top and b) side views of the most stable adsorption configuration of HCHO
     on P-doped CeO$_{2}$(111).}
    \label{formal}
    \end{center}
\end{figure}
%##############################

The adsorption of the HCHO molecule was systematically investigated by positioning it at various plausible adsorption sites on the P-doped CeO$_2$(111) surface.
Given that P$^{5+}$ is a highly charged cation relative to Ce, its presence introduces a Lewis acidic site on the CeO$_{2}$(111) surface. These sites are expected to strongly interact with the oxygen atom of the HCHO molecule, thereby enhancing adsorption strength and promoting molecular activation. Figure {\color{blue}\ref{formal}} shows the most energetically favorable adsorption configuration, in which the oxygen atom of the carbonyl group binds directly to the P dopant (d$_{P-O}$~=~1.57~\AA{}), while the carbon atom interacts with a surface oxygen (O$_{2c}$) with a C$-$O bond length of~1.37~\AA. This adsorption configuration yields a strong adsorption energy of -1.99 eV, which is significantly lower than that reported for HCHO adsorption on Mn-doped CeO$_2$(111) surfaces{\color{blue}\cite{SONG2022127985}}.

We also investigated the adsorption of HCHO on the pristine CeO$_2$(111) surface. In the most stable configuration, the carbon atom of HCHO binds to a surface O atom, while the oxygen atom of the carbonyl group coordinates to a surface Ce atom. The corresponding adsorption energy is -0.62 eV, indicating relatively weak adsorption. This value is in good agreement with the results reported by $Jiang ~et ~al.${\color{blue}\cite{WHXB20081115}} and is slightly lower than that obtained by $Meizan ~et ~al.${\color{blue}\cite{acsjpcc7b09276}}.

To gain insight into the effect of the adsorption on the CeO$_2$(111) surface, we calculated the work function ($\Phi$) of the studied systems using
$\Phi = E_{v} - E_{F}$,
where $E_{v}$ and $E_{F}$ denote the vacuum level and the Fermi level, respectively.

For the undoped CeO$_{2}$(111) surface, $\Phi$ was determined to be 6.46 eV, which decreased to 6.06 eV following gas adsorption. Notably, P$-$doping reduced the baseline work function to 5.59 eV, with further adsorption lowering it to 5.34 eV.
This systematic reduction in $\Phi$ reflects the enhanced ability of the P-doped surface to donate electron density toward adsorbed species
which reduces the activation barrier for charge transfer to adsorbed species.  Such enhanced electron availability directly supports the catalytic cycle by facilitating the activation of reaction intermediates at the surface, consistent with the adsorption energy and electronic structure findings discussed above.

\begin{figure}[h!]
   \begin{center}
       \includegraphics[angle=0,width=0.8\linewidth,clip]{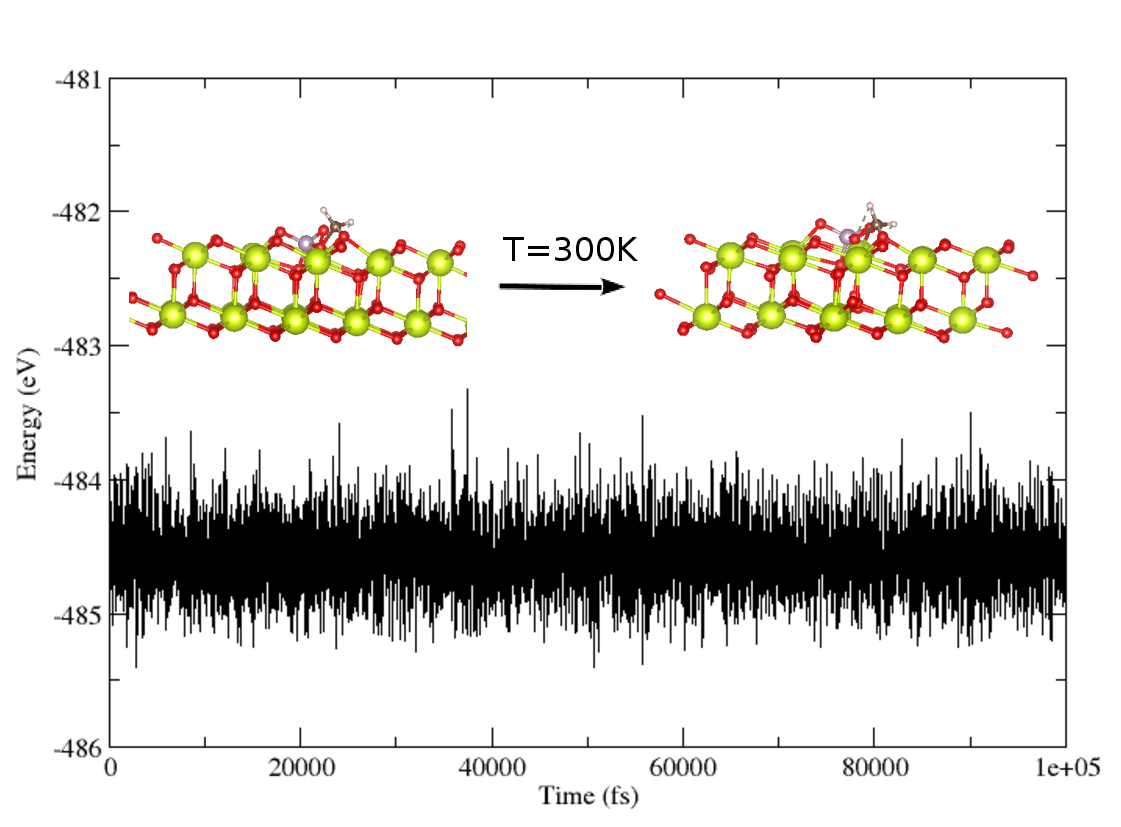}
\caption{Energy profile obtained from ab initio molecular dynamics (AIMD) simulations of the HCHO adsorbed on the P-doped CeO$_2$(111) surface.}
    \label{dat1}
   \end{center}
\end{figure}

To assess the thermal stability of the structure, we conducted ab initio molecular
dynamics (AIMD) simulations at 300 K for 100 $ps$. Under the $NVT$ ensemble,
regulated by a Nos\'e-Hoover thermostat, the simulation showed no significant energy
fluctuations or structural deformations (Fig.{\color{blue}\ref{dat1}}),
confirming the material's thermal stability at room temperature.

\begin{figure}[h!]
   \begin{center}
       \includegraphics[angle=0,width=0.6\linewidth,clip]{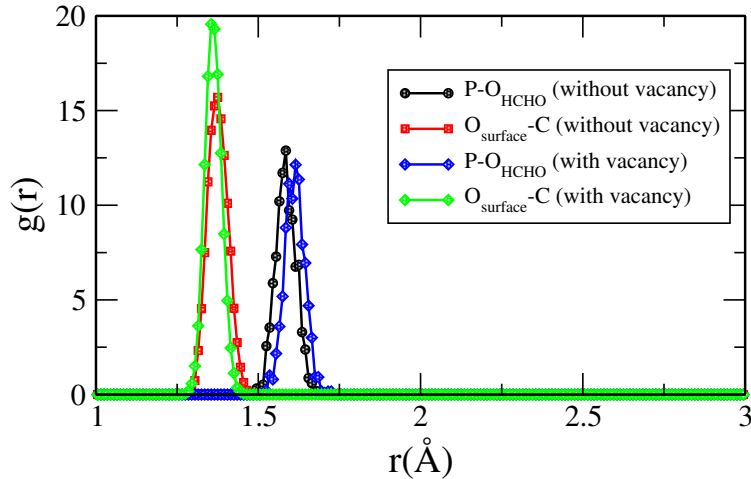}
\caption{
Radial distribution functions (RDFs) of the P$-$O${_{HCHO}}$ and O${_{surface}}$$-$C interatomic distances describing HCHO interactions with the P-doped CeO$_2$(111) surface, with and without oxygen vacancy.
}
    \label{rdf1}
   \end{center}
\end{figure}

To further quantify the structural stability and bonding characteristics at finite temperature, radial distribution functions (RDFs) were calculated from the AIMD trajectories at 300 K. The RDF analysis provides statistical information on the time-averaged distances between surface atoms and adsorbate species, allowing a quantitative assessment of the persistence of covalent interactions under thermal conditions.\\
As shown in Fig. {\color{blue}\ref{rdf1}}, the RDFs at 300 K for the P$-$O$_{HCHO}$ and O$_{surface}$$-$C pairs exhibit single pronounced first peaks located at 1.59 \AA~ and 1.38 \AA, respectively. These distances are consistent with covalent bonding and indicate that the adsorbed HCHO molecule remains strongly anchored to the surface throughout the simulation.

%##############################

%##############################

%##############################

%##############################

\subsection{HCHO and O$_{2}$ adsorption on the defective P-doped CeO$_{2}$(111) surface}
\begin{figure}[H]
   \begin{center}
       \includegraphics[angle=0,width=0.8\linewidth,clip]{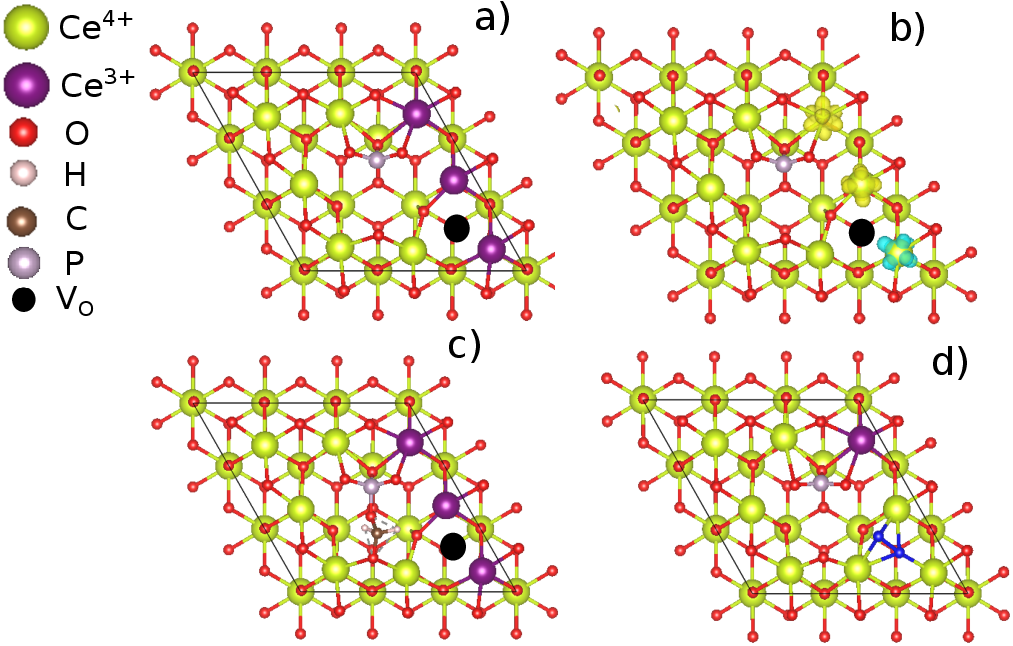}
    \caption{(a) Top view of the most stable defective P-doped CeO$_2$(111) surface. (b)The spin-density plot of the defective P-doped CeO$_2$(111) surface. The different colors of the volumetric data correspond to different spin projections.
(c) and (d) Optimized adsorption configurations of HCHO and O$_2$, respectively, on the defective P-doped CeO$_2$(111) surface. The O$_2$ molecule is highlighted in blue.}
    \label{defect}
    \end{center}
\end{figure}

Surface defects, particularly oxygen vacancies, play a crucial role in enhancing the photocatalytic and catalytic activity of CeO$_2$. In the present
work three classes of candidate sites were considered: oxygen positions in the immediate coordination shell of the phosphorus dopant, surface sites located at increasing distances from the P atom, and subsurface oxygen positions beneath the doped layer. For each candidate, full structural relaxation was performed and the resulting energy values were compared.
The site highlighted by the black circle in Fig.{\color{blue}\ref{defect}(a,b)}  consistently yielded the lowest formation energy across all tested configurations, confirming it as the most energetically favorable vacancy position.
This defect generates two excess electrons, which localize on neighboring cerium atoms and occupy the Ce 4$f$ orbitals, leading to the formation of two additional Ce$^{3+}$ ions with a Bader charge of +2.09$\lvert e \rvert$ (Fig.{\color{blue}\ref{defect}(a,b)}).

The oxygen vacancy formation energy ($E_f(V_O)$) calculated for the most stable configuration of the P-doped $\text{CeO}_2(111)$ surface is -0.64 eV. This value is more negative than that of the pristine CeO$_2$(111) surface (2.96 eV) and well below those previously reported for Sn- and Au-doped CeO$_2$(111) systems {\color{blue}\cite{celc.202100445, acsjpcc7b09276}}. Such a pronounced reduction signals that, under the oxygen-rich conditions employed throughout this work, the stoichiometric P/CeO$_2$(111) surface does not represent a thermodynamically stable configuration and is instead susceptible to spontaneous oxygen loss. The system should therefore be treated as a phosphorus-doped, oxygen-deficient CeO$_{2-x}$ phase. Analogous behavior is observed for several  dopants: on the (111) surface, Cu, Zn, Cd, Hg, Ag, Mo, and Os all give negative E$_{f}$(V$_{0}$) values, while on the more reactive (110) surface the same holds for Cr, Fe, Pd, Tc, Co, and Ru {\color{blue}\cite{D3NR05950B}}.
The negative formation energy can be understood in terms of the charge imbalance introduced when the pentavalent P$^{5+}$ occupies a Ce$^{4+}$ lattice site: the resulting local charge excess is partially relieved by the removal of a neighboring oxygen atom, which simultaneously drives the reduction of two adjacent Ce$^{4+}$ ions to Ce$^{3+}$. Under oxygen-rich conditions, both this charge compensation and the accompanying structural relaxation around the dopant site contribute to making vacancy formation intrinsically favorable.
Consequently, the P-doped CeO$_2$(111) surface is expected to exhibit a high concentration of oxygen vacancies under reaction conditions. These defects can act as active sites for adsorption and activation of reactant molecules, highlighting the crucial role of phosphorus doping in promoting defect formation and emphasizing the importance of investigating the catalytic mechanisms of formaldehyde oxidation on defective P-doped CeO$_2$(111) surfaces.

%##############################
%##############################
For the HCHO molecule, multiple adsorption sites and conﬁgurations are investigated on the defective P-doped CeO$_{2}$(111) surface. In Fig.{\color{blue}\ref{defect}c} we show the most favorable adsorption configuration. In this configuration, the oxygen atom of the carbonyl group coordinates with the P atom (1.58 \AA), while the carbon atom binds to a twofold-coordinated surface oxygen atom (O$_{2c}$) with a bond length of 1.37 \AA. The valence state of the P atom remains unchanged at $+5$ upon HCHO adsorption. The calculated adsorption energy of -2.65 eV reveals a strong chemisorptive
interaction between the formaldehyde molecule and the P-doped CeO$_{2}$(111) surface. This value is significantly larger in magnitude than those reported for ceria surfaces doped with noble metals or transition metals {\color{blue}\cite{acsjpcc7b09276,SONG2022127985,JING2022133599}}, highlighting the enhanced adsorption capability induced by the combined effects of phosphorus doping and oxygen vacancies.

\begin{figure}[H]
  \begin{center}
       \includegraphics[angle=0,width=0.8\linewidth,clip]{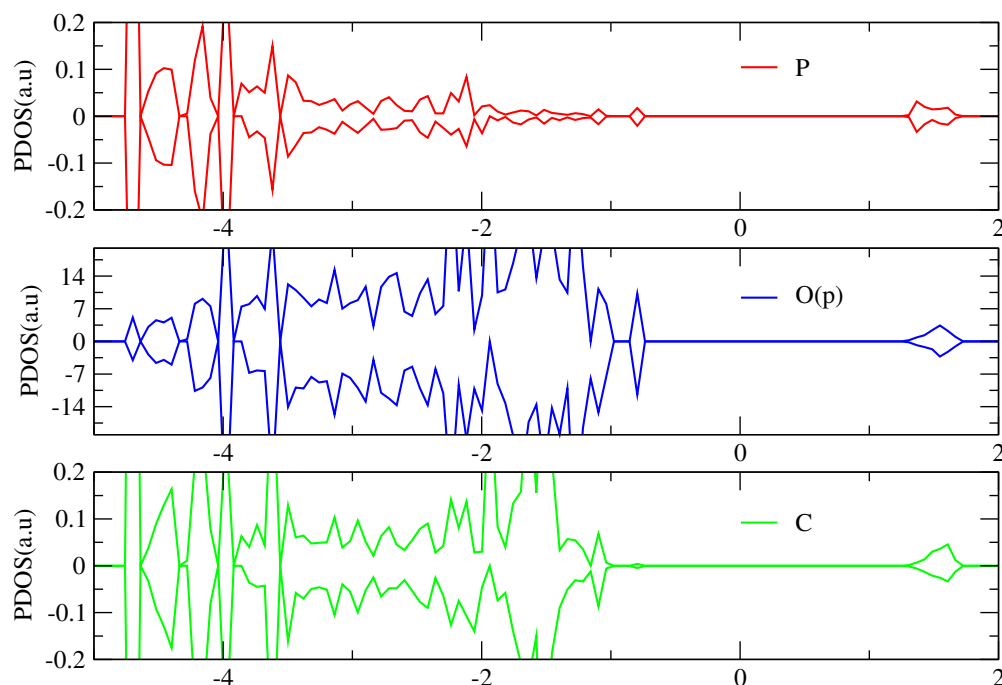}
     \caption{Partial density of states (PDOS) of the dopant (P) and the HCHO molecule's carbon and oxygen atoms.}
    \label{PDOSH}
    \end{center}
\end{figure}

The strong adsorption of HCHO can be rationalized by analyzing the partial density of states (PDOS) of the dopant atom and the carbon and oxygen atoms of the HCHO molecule. As shown in the PDOS plot (Fig.{\color{blue}\ref{PDOSH}}), the strong overlap between the filled p-orbitals of the dopant and the p states of the oxygen atom of the HCHO molecule indicates the formation of a strong covalent bond and significant orbital hybridization. This hybridization gives rise to new bonding states at lower energies, which stabilize the adsorbed configuration and account for the large adsorption energy.

While the calculated adsorption energy indicates a very strong interaction between HCHO and the defective P-doped CeO$_2$(111) surface, such strong adsorption may raise questions regarding the catalytic efficiency, since excessively strong binding can potentially lead to surface site blocking. In heterogeneous catalysis, an optimal balance between adsorption strength and catalytic activity is generally required, as described by the Sabatier principle. In the present case, the strong adsorption of HCHO is expected to facilitate the activation of the C$-$H and C$-$O bonds, which is a key step in the oxidation mechanism of formaldehyde.

%##############################
%##############################

\begin{figure}[h!]
   \begin{center}
\includegraphics[angle=0,width=0.48\linewidth,clip]{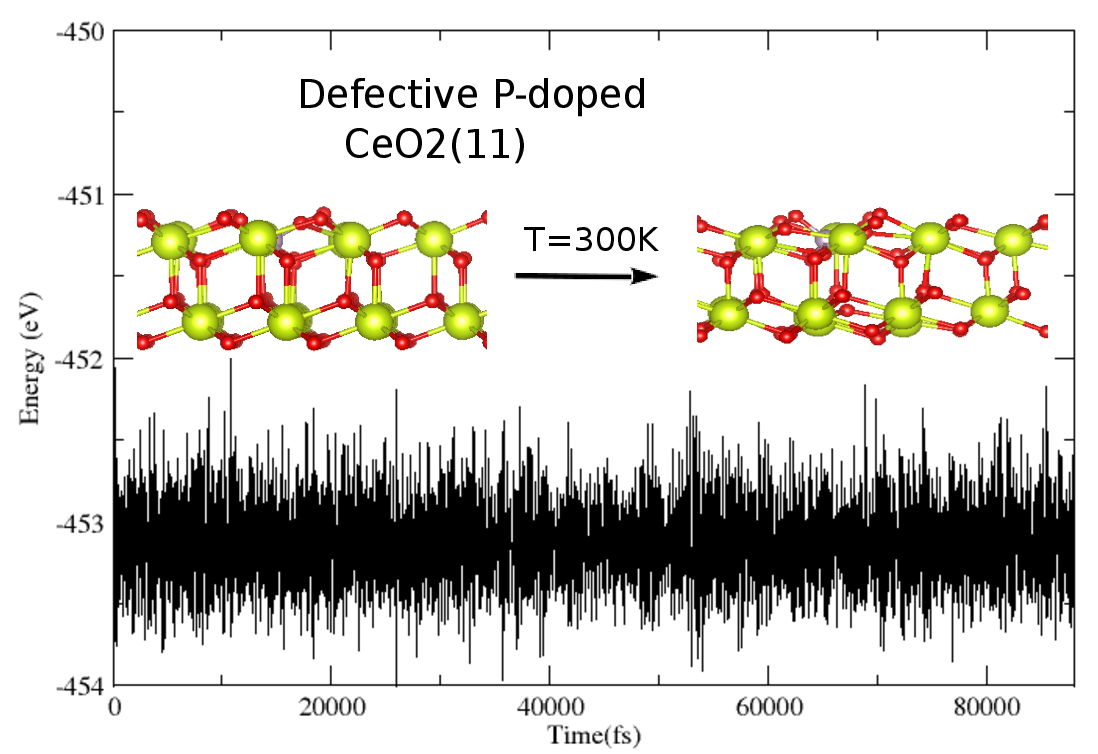}
      \includegraphics[angle=0,width=0.48\linewidth,clip]{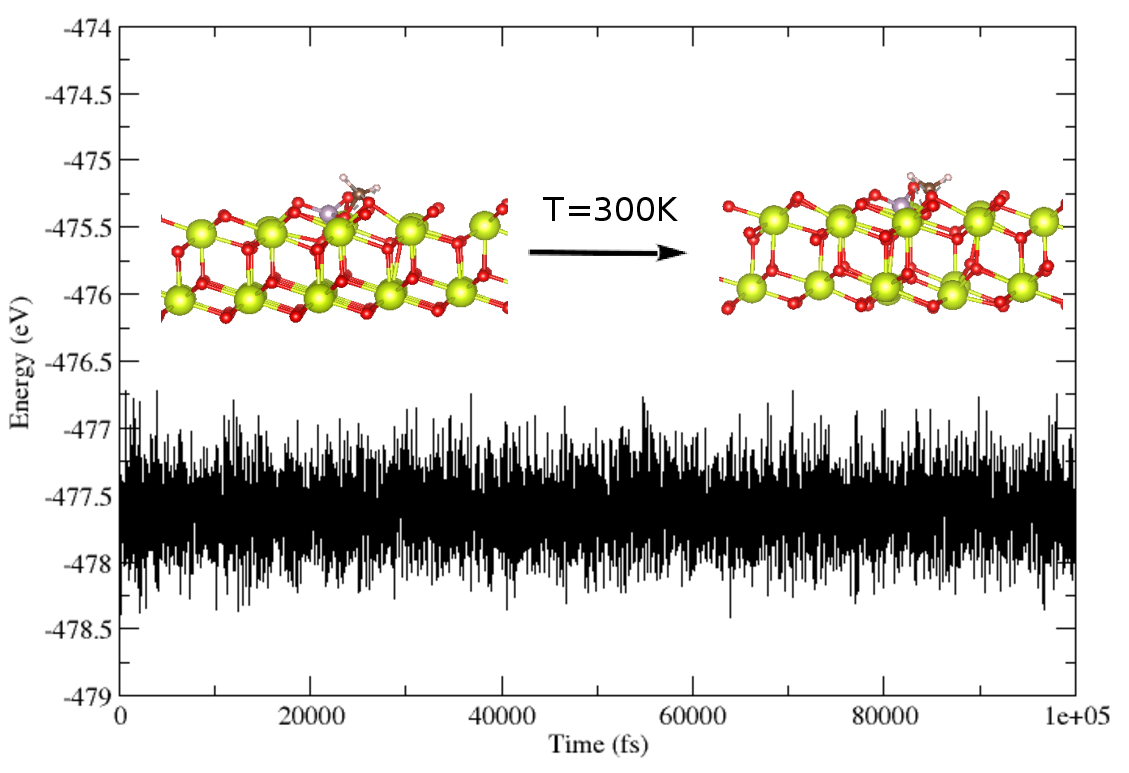}
  \caption{ Energy profiles obtained from ab initio molecular dynamics (AIMD) simulations of the defective P-doped CeO$_2$(111) surface with and without adsorbed HCHO.}
    \label{AIMD1}
  \end{center}
  \end{figure}

Fig.{\color{blue}\ref{AIMD1}} reports the energy evolution during the AIMD
simulations of the defective P-doped CeO$_{2}$(111) with and without HCHO adsorbed
molecule. This explorative study reveals a thermal stability of the two structures at 300 K. The adsorption energy of HCHO is found to be -2.43 eV at 300 K. The RDF analysis, presented in Fig. {\color{blue}\ref{rdf1}}, further confirms that the HCHO molecule remains strongly adsorbed on the surface, with the P-O$_{HCHO}$ and O$_{surface}$$-$C interatomic distances centered at approximately 1.62 \AA~ and 1.35 \AA, respectively. However, an adsorption energy of -1.32 eV is observed when molecular oxygen adsorbs
on the surface with T = 300 K, suggesting that HCHO presents stronger adsorption capability
than O$_{2}$. This value is lower than those of peroxide species in reduced CeO$_{2}$(111) surface{\color{blue}\cite{10.1021jp3016326}}.

%#$#############################
%##############################

The most stable adsorption configuration is shown in Fig.{\color{blue}\ref{defect}c}. In this geometry, one oxygen atom of the O$_2$ molecule occupies the oxygen vacancy, while the second oxygen atom interacts with a surface cerium atom. The excess electron initially localized on a neighboring Ce atom is transferred to the adsorbed O$_2$ molecule. As a consequence, the O$-$O bond length increases from 1.23 \AA~in the gas$-$phase O$_2$ molecule to 1.43 \AA~upon adsorption. This significant bond elongation indicates strong activation of the O$_2$ molecule, which facilitates its adsorption and subsequent dissociation. Bader charge analysis further confirms that the adsorbed O$_2$ species adopts a peroxide$-$like O$_2^{2-}$ configuration.

\subsection{HCHO and O$_{2}$ co-adsorbing on the defective P-doped CeO$_{2}$(111) surface}

Following the independent adsorption studies of HCHO and O$_2$ on the defective P-doped CeO$_2$(111) surface, we now investigate their co-adsorption behavior to elucidate the initial steps of the formaldehyde oxidation mechanism.
In this part two competitive adsorption sequences were explicitly compared: in Sequence A, HCHO adsorbs first on the defective P-CeO$_2$ surface followed by O$_2$ co-adsorption, while in Sequence B, O$_2$ occupies the oxygen vacancy first followed by HCHO adsorption on the partially reoxidized surface. The fully co-adsorbed state reached via Sequence A (HCHO + O$_2$ + P-CeO$_2$ + V$_O$) is found to be more stable than the analogous state reached via Sequence B (O$_2$ + HCHO + P-CeO$_2$ + V$_O$), with the same atomic composition, confirming that pre-adsorption of HCHO yields the thermodynamically favored configuration.

\begin{figure}[H]
   \begin{center}
       \includegraphics[angle=0,width=0.8\linewidth,clip]{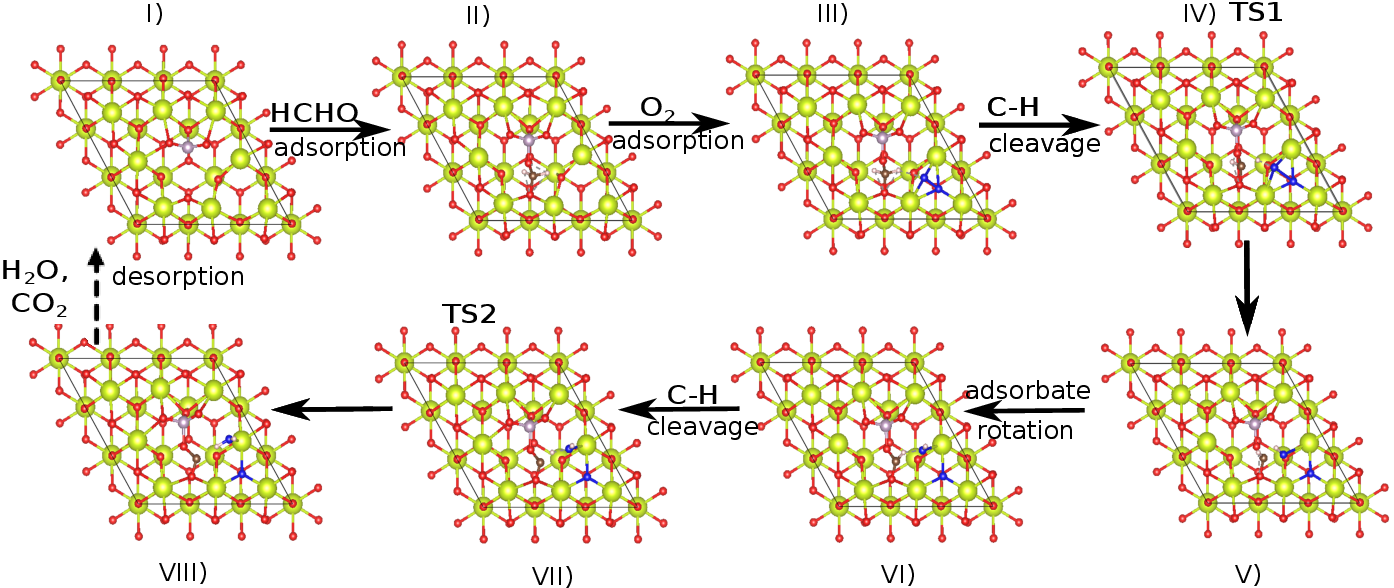}
       \includegraphics[angle=0,width=0.65\linewidth,clip]{deltae.eps}
     \caption{ Mechanism of formaldehyde oxidation on the defective P-doped
     CeO$_{2}$(111) surface.}
    \label{oscid}
    \end{center}
\end{figure}

Figure {\color{blue}\ref{oscid}} presents the energy profile along with the optimized structures of the reaction intermediates and transition states (TSs) involved in HCHO oxidation in the presence of O$_2$ on the defective P-doped CeO$_2$(111) surface.\\
When HCHO is pre-adsorbed on the P-doped CeO$_2$(111) surface, molecular oxygen subsequently adsorbs exothermically with an energy change of -1.77 eV, corresponding to state III. During this process, the O$-$O bond is significantly elongated to 1.48 \AA, indicating strong activation of the O$_2$ molecule, and a hydrogen bond is formed between O$_2$ and the molecule HCOH.

%##############################
\begin{figure}[H]
   \begin{center}
     \includegraphics[angle=0,width=0.8\linewidth,clip]{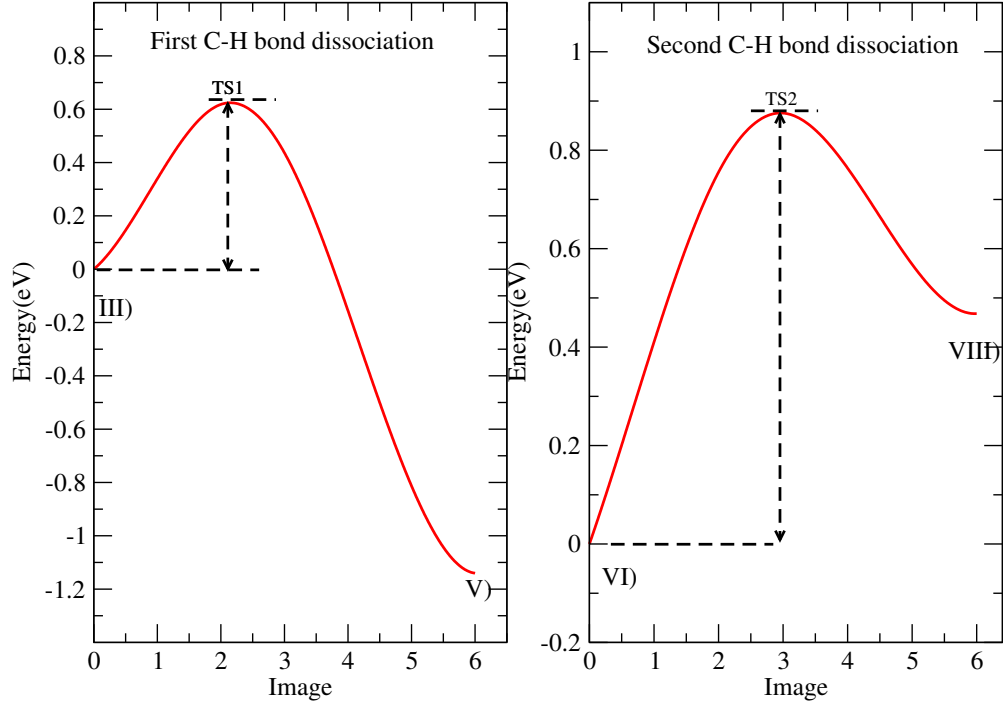}
     \caption{Activation barrier with the transition states for the first and the
     second C-H bond dissociations.}
    \label{neb}
    \end{center}
\end{figure}

%##############################
This adsorption process is accompanied by electron transfer from Ce$^{3+}$ to the adsorbed O$_2$ molecule, leading to the formation of a peroxide-like O$_2^{2-}$ species. The subsequent reaction step involves cleavage of a C–H bond in the adsorbed HCHO molecule, followed by hydrogen transfer to the activated O$_2^{2-}$ species. This step is exothermic, with a reaction energy of -1.67 eV, and proceeds via a transition state with an activation barrier of 0.62 eV (TS1, Fig.{\color{blue}\ref{neb}}). Notably, this barrier is significantly lower than those reported for the stoichiometric CeO$_2$(111) surface (1.71 eV){\color{blue}\cite{TENG201068}} and Mn-doped CeO$_2$(111){\color{blue}\cite{acs.jpcc.6b03218}}, and is comparable to that obtained for Au-doped CeO$_2$(111){\color{blue}\cite{acsjpcc7b09276}}.
This reaction results in cleavage of the O$-$O bond, leaving surface-bound OH$^{-}$ and CHO$^{-}$ species (state V). Following this step, the generated OH$^{-}$ and CHO$^{-}$ species undergo a reorientation with an associated energy change of 0.26 eV (state VI), leading to a configuration favorable for accepting the second hydrogen atom. The reaction then proceeds through the cleavage of the second C$-$H bond, with the released hydrogen atom transferred to the OH$^{-}$ species, resulting in the formation of H$_2$O and CO$_2$. This second C$-$H bond dissociation occurs via a transition state with an activation barrier of 0.87 eV (TS2, Fig.{\color{blue}\ref{neb}}), which is significantly lower than that reported for the stoichiometric CeO$_2$(111) surface{\color{blue}\cite{TENG201068}}, and requires an endothermic energy input of 0.46 eV (state VII). The generated CO$_{2}$ can easily desorb from the surface of the catalyst (state IX) by costing energy of 0.27 eV, creating an oxygen vacancy on the surface of the catalyst, COO$\rightarrow$CO$_{2}$(g)+V$_{O}$.
Finally, H$_2$O is desorbed from the surface of the catalyst (state X), H$_{2}$O$\rightarrow$H$_{2}$O(g),
costing an energy of 0.49 eV. Then the whole reaction cycle is
completed and the catalyst returns to the initial state.

%##############################
\begin{figure}[H]
   \begin{center}

       \includegraphics[angle=0,width=0.7\linewidth,clip]{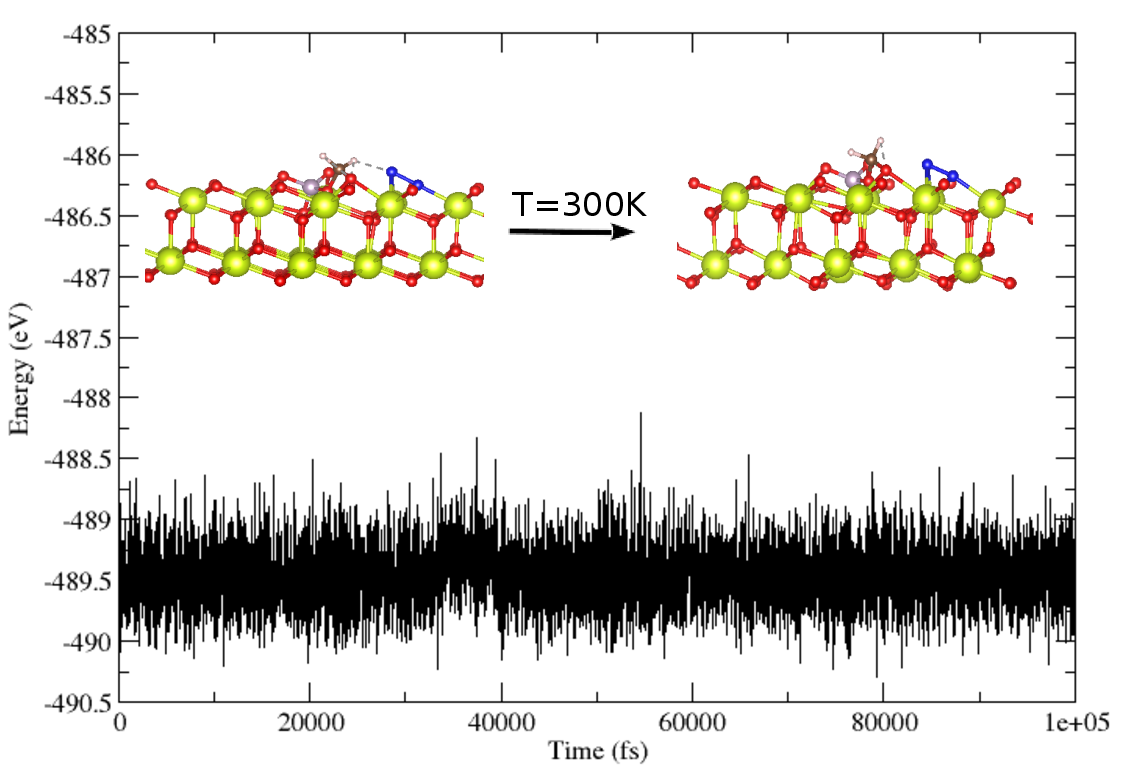}
    \caption{Potential energy evolution during the ab initio molecular dynamics (AIMD) simulation of the co-adsorbed HCHO and O$_2$ structure on the defective P-doped CeO$_2$(111) surface.}
    \label{AIMD11}
    \end{center}
\end{figure}

%##############################
For controllable catalysis to be successful, it is important that the catalysts
themselves maintain a good stability. To evaluate this, AIMD simulations were performed at room temperature for the co-adsorbed HCHO and O$_{2}$ system (State III), as illustrated in Fig.{\color{blue}\ref{AIMD11}}. The results indicate that thermal activation leads to a slight shortening of the O$-$O bond length to 1.44 \AA. This is followed by an OH reorientation during the 35$-$39 ps interval, resulting in the dissociation of a hydrogen bond. Apart from this reorientation, no significant structural distortion is observed in the overall system, indicating good thermal stability. At room temperature, when HCHO is adsorbed on the defective P-doped CeO$_2$(111) surface, O$_2$ adsorbs exothermically with an adsorption energy of -1.59 eV.

%##############################
%##############################
The time required for the desorption of CO$_{2}$ and H$_{2}$O molecules from the catalyst surface is crucial for its ability to maintain the oxidation reaction.
Thus, the desorption time ($\tau$) for these molecules in this specific reaction pathway can be calculated using the following equation:
\begin{equation}
 \tau = \frac{1}{A}exp(\frac{\Delta E_{des}}{k_{B}T})
\end{equation}
where k$_{B}$ is Boltzmann’s constant, T is the absolute temperature and A is the bond vibrational frequency taken as 10$^{13}$ Hz{\color{blue}\cite{ZHANG2025136006,debbichi16024016}}. $\Delta E_{des}$ is the desorption energy and is referred to as the activation energy of desorption. According to our calculations, the desorption time at 300 K of CO$_{2}$ and H$_{2}$O is $\sim$ 0.59 s. This value is similar to the desorption time of CO$_{2}$ and H$_{2}$O from N-doped Pd/MgAl{\color{blue}\cite{ZHANG2025136006}}, which suggests a weak interaction between the molecules and the catalyst surface which is desirable in catalysis.

These results provide important insight into the catalytic performance of the defective P-doped CeO$_2$(111) surface. Although the adsorption energy of HCHO is relatively large, which could in principle raise concerns about possible surface poisoning, the calculated reaction pathway and the short desorption time of the reaction products suggest that the catalytic cycle can proceed efficiently. In the present system, the adsorption energy of -2.65 eV observed on our defective P-doped surface is consistent with this same activation-driven chemisorption mechanism namely, the formation of a dioxymethylene-type intermediate rather than indicating passive site-blocking. While the calculated barriers for the first and second C-H bond cleavage steps (0.62 eV and 0.87 eV) exceed the most favorable values reported by  Song et al.{\color{blue}\cite{SONG2022127985}}, they remain substantially lower than those obtained for unreactive stoichiometric surfaces (1.15-1.32 eV){\color{blue}\cite{TENG201068,SONG2022127985}}. This places our results in line with active catalytic turnover at moderate temperatures, rather than surface deactivation.
At the same time, the weak interaction of the final products (CO$_2$ and H$_2$O) with the surface, reflected by their short desorption time, prevents surface blocking and ensures the regeneration of active sites. Therefore, the defective P-doped CeO$_2$(111) surface appears to achieve a favorable compromise between strong reactant activation and efficient product removal, which is a key requirement for efficient catalytic performance in formaldehyde oxidation.

\section{Conclusion}
In this work, using DFT simulations we evaluate the synergistic effects of phosphorus doping and oxygen vacancies on the oxidation of formaldehyde on CeO$_2$(111) surface. The results show that the doped P atom adopts a pentavalent state (P$^{5+}$) and induces charge redistribution toward neighboring atoms, leading to the reduction of a Ce$^{4+}$ ion to Ce$^{3+}$.
Our results demonstrate that phosphorus doping markedly enhances the adsorption energy by strengthening the interaction between formaldehyde and the phosphorus-doped CeO$_{2}$(111) surface.
Under the computational conditions employed in this study, P-doped CeO$_{2}$(111) shows improved performance relative to both undoped ceria and ceria doped with noble (Au) or transition (Mn) metals.
Furthermore, the presence of P strongly decreases the oxygen vacancy formation energy of ceria. Compared to the stoichiometric CeO$_{2}$(111) surface, this mechanism exhibits substantially reduced barriers for C$-$H bond cleavage during formaldehyde oxidation, indicating that doping with P strongly activates this oxidation.
Finite-temperature AIMD simulations reveal no significant energy fluctuations or structural deformations at room temperature for the adsorbed systems. Finally, the relatively short desorption time ($\sim$ 0.59 s at 300 K) demonstrates that CO$_{2}$ and H$_{2}$O interact only weakly with the catalyst, facilitating rapid release from the surface. These findings demonstrate that phosphorus-doped ceria is a new, highly effective way for the oxidation of formaldehyde.

\section*{Conflicts of interest} There are no Conflicts to declare.

\section*{Acknowledgments} The study was  supported by the France 2030 framework under the PEPR DIADEM (project ANR-23-PEXD-0007)


\begin{thebibliography}{10}

\bibitem{Chen2017}
Hui Chen, Jiabo Hu, Guo-Dong Li, Qian Gao, Cundi Wei, and Xiaoxin Zou.
\newblock Porous ga–in bimetallic oxide nanofibers with controllable
  structures for ultrasensitive and selective detection of formaldehyde.
\newblock {\em ACS Appl. Mater. Interfaces.}, 9:4692, 2017.

\bibitem{Liu2023}
Ru~Liu, Min Liang, Jianfeng Xu, Yuhui Sun, Ling Long, Liming Zhu, Bin Lv, Bohan
  Yang, and Yonghao Ni.
\newblock Preparation of a novel formaldehyde-free impregnated decorative paper
  containing mno2 nanoparticles for highly efficient formaldehyde removal.
\newblock {\em ACS Appl. Mater. Interfaces.}, 15(29):34941--34955, 2023.

\bibitem{He2021}
Taohong He, Yu~Zhou, Danni Ding, and Shaopeng Rong.
\newblock Engineering manganese defects in mn3o4 for catalytic oxidation of
  carcinogenic formaldehyde.
\newblock {\em ACS Appl. Mater. Interfaces.}, 13(25):29664--29675, 2021.

\bibitem{LIU2023110080}
Ningrui Liu, Lin Fang, Wei Liu, Haidong Kan, Zhuohui Zhao, Furong Deng, Chen
  Huang, Bin Zhao, Xiangang Zeng, Yuexia Sun, Hua Qian, Jinhan Mo, Chanjuan
  Sun, Jianguo Guo, Xiaohong Zheng, Zhongming Bu, Louise~B. Weschler, and
  Yinping Zhang.
\newblock Health effects of exposure to indoor formaldehyde in civil buildings:
  A systematic review and meta-analysis on the literature in the past 40 years.
\newblock {\em Building and Environment}, 233:110080, 2023.

\bibitem{ZHOU2025137831}
Xiaojun Zhou, Wenlong Li, Xinke Wang, and Yingying Wang.
\newblock Health risk assessment of indoor formaldehyde exposure across chinese
  residences: Effects of building material grades.
\newblock {\em Journal of Hazardous Materials}, 490:137831, 2025.

\bibitem{QI2021106293}
Qiuping Qi, Wenrui Zhang, Yanshi Zhang, Guangming Bai, Shiwei Wang, and Peng
  Liang.
\newblock Formaldehyde oxidation at room temperature over layered {MnO2}.
\newblock {\em Catal. Commun.}, 153:106293, 2021.

\bibitem{DENG2019540}
Jianlin Deng, Weiyu Song, Lulu Chen, Lu~Wang, Meizan Jing, Yu~Ren, Zhen Zhao,
  and Jian Liu.
\newblock The effect of oxygen vacancies and water on hcho catalytic oxidation
  over co3o4 catalyst: A combination of density functional theory and
  microkinetic study.
\newblock {\em Chem. Eng. J.}, 355:540--550, 2019.

\bibitem{TENG201068}
Bo-Tao Teng, Shi-Yu Jiang, Zong-Xian Yang, Meng-Fei Luo, and You-Zhao Lan.
\newblock A density functional theory study of formaldehyde adsorption and
  oxidation on {CeO2(111)} surface.
\newblock {\em Surf. Sci.}, 604:68--78, 2010.

\bibitem{BREYSSE197354}
Michéle Breysse, Michelle Guenin, Bernard Claudel, and Jean Veron.
\newblock Catalysis of carbon monoxide oxidation by cerium dioxide: Ii.
  microcalorimetric investigation of adsorption and catalysis.
\newblock {\em Journal of Catalysis}, 28(1):54--62, 1973.

\bibitem{TANG2008115}
Xingfu Tang, Junli Chen, Xiumin Huang, Yide Xu, and Wenjie Shen.
\newblock {Pt/MnOx–CeO2} catalysts for the complete oxidation of formaldehyde
  at ambient temperature.
\newblock {\em Appl. Catal. B: Environ.}, 81:115--121, 2008.

\bibitem{LIU2012467}
Baocang Liu, Changyan Li, Yifei Zhang, Yang Liu, Wenting Hu, Qin Wang, Li~Han,
  and Jun Zhang.
\newblock Investigation of catalytic mechanism of formaldehyde oxidation over
  three-dimensionally ordered macroporous {Au/CeO2} catalyst.
\newblock {\em Appl. Catal. B: Environ.}, 111-112:467--475, 2012.

\bibitem{D0CY01894E}
Xiaoxiao Qin, Xueyan Chen, Min Chen, Jianghao Zhang, Hong He, and Changbin
  Zhang.
\newblock Highly efficient {Ru/CeO2} catalysts for formaldehyde oxidation at
  low temperature and the mechanistic study.
\newblock {\em Catal. Sci. Technol.}, 11:1914--1921, 2021.

\bibitem{ZHANG2020121693}
Yajing Zhang, Yuqin Xiao, Lu~Li, Keke Song, Xiaoxu Wang, Chutian Wang, Xiaodong
  Jian, Chunlin Ji, and Ping Qian.
\newblock Formaldehyde oxidation on {Co}-doped reduced {CeO2(111)}:
  First-principles calculations.
\newblock {\em Surf. Sci.}, 701:121693, 2020.

\bibitem{NING2024133282}
Deyang Ning, Junshan Zhang, Arun Murali, Yuanpei Lan, Chaoyi Chen, Shaoyan
  Yang, Wei Zhang, and Junqi Li.
\newblock Advancements in organic pollutant remediation: The role of
  nitrogen-doped {rGO-CeO2} in photocatalytic efficiency enhancement.
\newblock {\em Colloids Surf. A: Physicochem. Eng. Asp.}, 685:133282, 2024.

\bibitem{acsomega.3c01305}
Jihun Paick, Seunghee Hong, Ji-Young Bae, Jy-Young Jyoung, Eun-Sook Lee, and
  Doohwan Lee.
\newblock Effective atomic {N} doping on {CeO2} nanoparticles by
  environmentally benign urea thermolysis and its significant effects on the
  scavenging of reactive oxygen radicals.
\newblock {\em ACS Omega}, 8:22646--22655, 2023.

\bibitem{D0NJ03270K}
Dung~Van Dao, Thuy T.~D. Nguyen, Thanh~Duc Le, Jeong-Mo Yoon, In-Hwan Lee, and
  Yeon-Tae Yu.
\newblock {Pd} supported {N}-doped {CeO2} as an efficient hydrogen oxidation
  reaction catalyst in {PEMFC}.
\newblock {\em New J. Chem.}, 44:17203--17207, 2020.

\bibitem{HAN2022132154}
Zhipeng Han, Zhuang Li, Yi~Li, Denghui Shang, Liangbo Xie, Yueqin Lv, Sihui
  Zhan, and Wenping Hu.
\newblock Enhanced electron transfer and hydrogen peroxide activation capacity
  with {N}, {P}-codoped carbon encapsulated {CeO2} in heterogeneous
  electro-fenton process.
\newblock {\em Chemosphere}, 287:132154, 2022.

\bibitem{YOU201847}
Yanchen You, Chuanning Shi, Huazhen Chang, Lei Guo, Liwen Xu, and Junhua Li.
\newblock The promoting effects of amorphous {CePO4} species on
  phosphorus-doped {CeO2/TiO2} catalysts for selective catalytic reduction of
  {NOx} by {NH3}.
\newblock {\em Molecular Catalysis}, 453:47--54, 2018.

\bibitem{cm504734a}
Carlos Sotelo-Vazquez, Nuruzzaman Noor, Andreas Kafizas, Raul Quesada-Cabrera,
  David~O. Scanlon, Alaric Taylor, James~R. Durrant, and Ivan~P. Parkin.
\newblock Multifunctional {P}-doped {TiO2} films: A new approach to
  self-cleaning, transparent conducting oxide materials.
\newblock {\em Chemistry of Materials}, 27:3234--3242, 2015.

\bibitem{acs.est.2c00942}
Kai Shen, Biao Gao, Hangqi Xia, Wei Deng, Jiaorong Yan, Xiaohan Guo, Yanglong
  Guo, Xingyi Wang, Wangcheng Zhan, and Qiguang Dai.
\newblock Oxy-anionic doping: A new strategy for improving selectivity of
  {Ru/CeO2} with synergetic versatility and thermal stability for catalytic
  oxidation of chlorinated volatile organic compounds.
\newblock {\em Environ. Sci. Technol.}, 56:8854--8863, 2022.

\bibitem{acs.jpcc.0c03649}
Sara Navarro-Ja{\'e}n, Luis~F. Bobadilla, Francisca Romero-Sarria, Oscar~H.
  Laguna, Nicolas Bion, and Jos{\'e}~A. Odriozola.
\newblock Evaluation of the oxygen mobility in {CePO4}-supported catalysts:
  Mechanistic implications on the water–gas shift reaction.
\newblock {\em J. Phys. Chem. C}, 124:16391--16401, 2020.

\bibitem{acs.jpcc.5b07734}
Weiyuan Yao, Yue Liu, Xiaoqiang Wang, Xiaole Weng, Haiqiang Wang, and Zhongbiao
  Wu.
\newblock The superior performance of sol-gel made {Ce-O-P} catalyst for
  selective catalytic reduction of {NO} with {NH3}.
\newblock {\em J. Phys. Chem. C}, 120:221--229, 2016.

\bibitem{GUPTA2022100947}
S.K. Gupta, C.S. Datrik, B.~Modak, J.~Prakash, A.K. Debnath, P.~Modak, and
  K.~Sudarshan.
\newblock Synergistic effect of doping and defect in achieving white light
  emission and oxygen reduction catalysis in {Ce1-xSmxPO4}.
\newblock {\em Mater. Today Chem.}, 25:100947, 2022.

\bibitem{C8TA04603D}
Sara Navarro-Jaén, Miguel~Ángel Centeno, Oscar~Hernando Laguna, and
  José~Antonio Odriozola.
\newblock {Pt/CePO4} catalysts for the wgs reaction: influence of the
  water-supplier role of the support on the catalytic performance.
\newblock {\em J. Mater. Chem. A}, 6:17001--17010, 2018.

\bibitem{acs.est.4c04436}
Qiguang Dai, Ronghua Xu, Hangqi Xia, Boyuan Qiao, Qiang Niu, Li~Wang, Aiyong
  Wang, Yun Guo, Yanglong Guo, Wei Wang, and Wangcheng Zhan.
\newblock Catalytic hydrolysis–oxidation of halogenated methanes over phase-
  and defect-engineered cepo4: Halogenated byproduct-free and stable
  elimination.
\newblock {\em Environ. Sci. Technol.}, 58:13562--13573, 2024.

\bibitem{GONG2020144314}
Kaili Gong, Keqing Zhou, and Bin Yu.
\newblock Superior thermal and fire safety performances of epoxy-based
  composites with phosphorus-doped cerium oxide nanosheets.
\newblock {\em Appl. Sur. Sci.}, 504:144314, 2020.

\bibitem{KRESSE1996}
G.~Kresse and J.~Furthm\"uller.
\newblock Efficient iterative schemes for ab initio total-energy calculations
  using a plane-wave basis set.
\newblock {\em Phys. Rev. B}, 54:11169, 1996.

\bibitem{GGA1996}
John~P. Perdew, Kieron Burke, and Matthias Ernzerhof.
\newblock Generalized gradient approximation made simple.
\newblock {\em Phys. Rev. Lett.}, 77:3865, 1996.

\bibitem{PhysRevB.57.1505}
S.~L. Dudarev, G.~A. Botton, S.~Y. Savrasov, C.~J. Humphreys, and A.~P. Sutton.
\newblock Electron-energy-loss spectra and the structural stability of nickel
  oxide: {A}n {LSDA+U} study.
\newblock {\em Phys. Rev. B}, 57:1505--1509, 1998.

\bibitem{D5CP01283J}
C.~Hachemi, H.~Dib, M.~Debbichi, M.~Badawi, C.~Eads, M.~Ibrahim, S.~Loridant,
  J.~Knudsen, H.~Kaper, and L.~Cardenas.
\newblock Persistence of {Ce}$^{3+}$ species on the surface of ceria during
  redox cycling: a modulated chemical excitation investigation.
\newblock {\em Phys. Chem. Chem. Phys.}, 27:12069--12079, 2025.

\bibitem{Grimme10}
Stefan Grimme, Jens Antony, Stephan Ehrlich, and Helge Krieg.
\newblock A consistent and accurate ab initio parametrization of density
  functional dispersion correction ({DFT-D}) for the 94 elements {H-Pu}.
\newblock {\em J. Chem. Phys.}, 132(15):154104, 2010.

\bibitem{Bader}
W~Tang, E~Sanville, and G~Henkelman.
\newblock A grid-based bader analysis algorithm without lattice bias.
\newblock {\em J. Phys.: Condens. Matter}, 21:084204, 2009.

\bibitem{neb}
Graeme Henkelman, Blas~P. Uberuaga, and Hannes Jónsson.
\newblock A climbing image nudged elastic band method for finding saddle points
  and minimum energy paths.
\newblock {\em J. Chem. Phys.}, 113(22):9901--9904, 2000.

\bibitem{Nose1984}
Shuichi Nosé.
\newblock A unified formulation of the constant temperature molecular dynamics
  methods.
\newblock {\em The Journal of Chemical Physics}, 81(1):511--519, 07 1984.

\bibitem{Hoover1985}
William~G. Hoover.
\newblock Canonical dynamics: Equilibrium phase-space distributions.
\newblock {\em Phys. Rev. A}, 31:1695--1697, Mar 1985.

\bibitem{Ayadi2022}
Tarek Ayadi, Sébastien Lebègue, and Michael Badawi.
\newblock Ab initio molecular dynamics investigation of the co-adsorption of
  iodine species with {CO} and {H2O} in silver-exchanged chabazite.
\newblock {\em Phys. Chem. Chem. Phys.}, 24:24992--24998, 2022.

\bibitem{PhysRevMaterials7065403}
Bianca Baldassarri, Jiangang He, Xin Qian, Emanuela Mastronardo, Sean
  Griesemer, Sossina~M. Haile, and Christopher Wolverton.
\newblock Accuracy of dft computed oxygen-vacancy formation energies and
  high-throughput search of solar thermochemical water-splitting compounds.
\newblock {\em Phys. Rev. Mater.}, 7:065403, 2023.

\bibitem{10.10631.1329672}
Graeme Henkelman, Blas~P. Uberuaga, and Hannes Jónsson.
\newblock A climbing image nudged elastic band method for finding saddle points
  and minimum energy paths.
\newblock {\em J. Chem. Phys.}, 113(22):9901--9904, 2000.

\bibitem{chem.201502170}
Surajit Some, Iman Shackery, Sun~Jun Kim, and Seong~Chan Jun.
\newblock Phosphorus-doped graphene oxide layer as a highly efficient flame
  retardant.
\newblock {\em Chemistry – A European Journal}, 21:15480--15485, 2015.

\bibitem{LIU2024119544}
Zhao Liu, Hongyang Ma, Charles~C. Sorrell, Pramod Koshy, Biao Wang, and Judy~N.
  Hart.
\newblock Enhancement of light absorption and oxygen vacancy formation in
  {CeO2} by transition metal doping: {A} {DFT} study.
\newblock {\em Appl. Catal. A: Gen.}, 670:119544, 2024.

\bibitem{D3NR05950B}
Ning Xu, Liangliang Xu, Yue Wang, Wen Liu, Wenwu Xu, Xiaojuan Hu, and
  Zhong-Kang Han.
\newblock Unraveling the formation of oxygen vacancies on the surface of
  transition metal-doped ceria utilizing artificial intelligence.
\newblock {\em Nanoscale}, 16:9853--9860, 2024.

\bibitem{acs.chemrev.5b00603}
Tiziano Montini, Michele Melchionna, Matteo Monai, and Paolo Fornasiero.
\newblock Fundamentals and catalytic applications of ceo2-based materials.
\newblock {\em Chemical Reviews}, 116:5987--6041, 2016.

\bibitem{acsanm3c02656}
Songlin Li, Min Zhang, Youqiang Dong, Jie Gao, Pengfei Cheng, and Hai Wang.
\newblock Density functional theory study of {P-D}oped {Co3O4(111)} facets for
  {HCHO} adsorption: Implications for metal oxide semiconductor gas sensors.
\newblock {\em ACS Appl. Nano Mater.}, 6(19):17501--17511, 2023.

\bibitem{SONG2022127985}
Weiyu Song, Lulu Chen, Lei Wan, Meizan Jing, and Zhi Li.
\newblock {T}he influence of doping amount on the catalytic oxidation of
  formaldehyde by {Mn-CeO2} mixed oxide catalyst: {A} combination of {DFT} and
  microkinetic study.
\newblock {\em J. Hazard. Mater.}, 425:127985, 2022.

\bibitem{WHXB20081115}
Shi-Yu. JIANG, Bo-Tao TENG, Ji-Qing. LU, Xue-Song LIU, Pei-Fang YANG, Fei-Yong
  YANG, and LUO Meng-Fei.
\newblock A density functional theory study of formaldehyde adsorption on
  {CeO2(111)} surface.
\newblock {\em Acta Physico-Chimica Sinica}, 24:2025, 2008.

\bibitem{acsjpcc7b09276}
Meizan Jing, Weiyu Song, Lulu Chen, Sicong Ma, Jianlin Deng, Huiling Zheng,
  Yongfeng Li, Jian Liu, and Zhen Zhao.
\newblock Density functional theory study of the formaldehyde catalytic
  oxidation mechanism on a {Au}-doped {CeO2(111)} surface.
\newblock {\em J. Phys. Chem. C}, 122:438--448, 2018.

\bibitem{celc.202100445}
Shunlian Ning, Zhiwei Guo, Jigang Wang, Shaobin Huang, Shaowei Chen, and
  Xiongwu Kang.
\newblock {Sn}-doped {CeO2} nanorods as high-performance electrocatalysts for
  {CO2} reduction to formate.
\newblock {\em ChemElectroChem}, 8:2680--2685, 2021.

\bibitem{JING2022133599}
Meizan Jing, Weiyu Song, Yongfeng Li, Zhen Zhao, Jian Liu, and Graeme
  Henkelman.
\newblock Theoretical study of structure sensitivity on {Au} doped {CeO2}
  surfaces for formaldehyde oxidation: The effect of crystal planes and {Au}
  doping.
\newblock {\em Chem. Eng. J.}, 433:133599, 2022.

\bibitem{10.1021jp3016326}
Yun Zhao, Bo-Tao Teng, Xiao-Dong Wen, Yue Zhao, Qiao-Ping Chen, Lei-Hong Zhao,
  and Meng-Fei Luo.
\newblock Superoxide and peroxide species on {CeO2(111)}, and their oxidation
  roles.
\newblock {\em J. Phys. Chem. C}, 116:15986--15991, 2012.

\bibitem{acs.jpcc.6b03218}
Hairong Wu, Sicong Ma, Weiyu Song, and Emiel J.~M. Hensen.
\newblock Density functional theory study of the mechanism of formaldehyde
  oxidation on {Mn-D}oped ceria.
\newblock {\em J. Phys. Chem. C}, 120:13071--13077, 2016.

\bibitem{ZHANG2025136006}
Wei Zhang, Mayi Zhou, Zehong Li, Zhaohui Chen, and Shuai Chen.
\newblock Mechanism of formaldehyde oxidation on the surface of {N}-doped
  {Pd/MgAl} catalysts: {A} combination of density functional theory and
  microkinetic study.
\newblock {\em Fuel}, 402:136006, 2025.

\bibitem{debbichi16024016}
M.~Debbichi, A.~Mallah, M.~Houcine Dhaou, and S.~Leb\`egue.
\newblock First-principles study of monolayer penta-{CoS2} as a promising anode
  material for li/na-ion batteries.
\newblock {\em Phys. Rev. Appl.}, 16:024016, 2021.

\end{thebibliography}
\end{document}